\documentstyle[prb,aps,twocolumn,epsf,floats]{revtex}
\def\subequations{\begin{mathletters}}
\def\endsubequations{\end{mathletters}}
\begin{document}
\draft 
\title{Structure and Magnetization of Two-Dimensional Vortex Arrays in the
Presence of Periodic Pinning}
\author{Toby Joseph \cite{toby} and Chandan Dasgupta \cite{chandan}}
\address{Centre for Condensed Matter Theory, Department of Physics,
Indian Institute of Science,
Bangalore 560012, India\\
and\\
Condensed Matter Theory Unit, JNCASR, Bangalore 560064, India}

\maketitle
\begin{abstract}
Ground-state properties of a two-dimensional system of superconducting vortices 
in the presence of a periodic array of strong pinning centers  are studied
analytically and numerically. The ground states of the vortex system at
different filling ratios are obtained using a simple geometric argument under
the assumption that the penetration depth is much smaller than the spacing 
of the pin lattice. The results of this calculation are confirmed by
numerical studies in which simulated annealing is used to locate the ground
states of the vortex system. The zero-temperature equilibrium magnetization
as a function of the applied field is obtained by numerically calculating the
energy of the ground state for a large number of closely spaced filling
ratios. The results show interesting commensurability effects 
such as plateaus in the $B-H$ diagram at simple fractional filling ratios. 

\end{abstract}
\section{Introduction}
\label{intro}
In the mixed phase of type-II superconductors, magnetic flux penetrates the
sample in the form of quantized vortex lines~\cite{tink}. The amount of flux
carried by each vortex line is equal to the basic flux quantum
$\Phi_0 = hc/(2e) = 2.07 \times 10^{-7}$G.cm$^2$. 
These vortex lines form a special physical system
known as ``vortex matter''. In the absence of any pinning sites in the 
material, the vortex lines form a triangular lattice known as the Abrikosov
lattice~\cite{abri}.

Equilibrium and transport properties of the mixed phase of type-II 
superconductors are strongly affected by the presence of pinning centers,
either intrinsic to the system or artificially generated. Understanding the
effects of pinning in these systems is very important for practical
applications because the presence of pinning strongly influences the value
of the critical current in the mixed phase.

In recent years, a variety of nanofabrication techniques have been used to
create periodic arrays of pinning centers in thin-film superconductors
~\cite{exp1,exp2,exp3,exp4,exp5,ion,elect,dot1,dot2,dot3,dot4}. Such arrays may 
consist of microholes (``antidots'')~\cite{exp1,exp2,exp3,exp4,exp5}, 
defects produced
by the bombardment of ion~\cite{ion} or electron~\cite{elect} beams, or
magnetic dots~\cite{dot1,dot2,dot3,dot4}. These pinning centers are ``strong''
in the sense that each pinning site can trap one or more vortices at low
temperatures. The effects of periodic pinning depend strongly on the
relative values of $B_\phi$ and $B$, where $B_\phi = \rho_p \Phi_0$ ($\rho_p$
is the areal density of the pinning centers) is the so-called ``matching 
field'', and $B$ is the magnetic induction that determines the areal density
$\rho_0$ of vortices ($\rho_0=B/\Phi_0$). The filling ratio, $n$, defined
as $n \equiv B/B_\phi$, measures the commensurability of the vortex 
system with the underlying pin lattice. The interplay between the lattice
constant of the pin array (determined by $B_\phi$) and the intervortex
separation (determined by $B$) can lead to a variety of interesting effects
in such systems.

Some of these effects have been observed in recent experiments. Imaging 
experiments using various techniques such as Lorentz microscopy~\cite{ion}
and scanning Hall probe microscopy~\cite{exp3,exp5} have shown the formation of
ordered structures of the vortex system at low temperatures for 
commensurate values of $n$. Magnetization measurements~\cite{exp1,exp2,exp4} in
the irreversible (vortex solid) regime have demonstrated the occurrence of 
anomalies at certain harmonics of $B_\phi$. The effectiveness of pinning at
integral values of $n$ has been found~\cite{exp4,dot1,dot2,dot3,dot4} to produce
regularly spaced sharp minima in the resistivity versus field curve. A 
pinning-induced reconfiguration of the vortex lattice has been 
observed~\cite{dot4} in a thin-film superconductor with a rectangular array 
of magnetic dots. Some of these effects have been studied theoretically,
using analytic~\cite{th1} and 
numerical~\cite{simul1,simul1a,simul1b,trigs,cdotv,simul2} methods.
Experimental realizations of a system of interacting ``particles'' in the
presence of an external periodic potential are also obtained in colloidal
suspensions in interfering laser fields~\cite{laser}, and in periodic
arrays of optical traps~\cite{traps}.

In this paper, we have used analytic and numerical methods to analyze the
zero-temperature structure of vortex arrays in the presence of periodic
pinning. We have also carried out a numerical study of the zero-temperature
{\it equilibrium} magnetization of a superconducting film with a square
array of pinning centers as a function of the applied field. In 
section~\ref{nlt1}, we consider the ground states of a vortex system 
in a square array of pinning centers 
for fillings less than unity. We look at a class of structures that are 
Bravais lattices with one vortex per basis if the filling $n$ is of the 
form $1/q$, and with $p$ vortices per basis if $n=p/q$ ($p$ and $q$ are
integers greater than unity, with $p < q$). The structure with the lowest
energy in this class can be obtained rather easily. We find that the
``ground-state'' structure obtained this way matches those obtained from
experiments~\cite{ion} and simulations~\cite{simul1,trigs} for a large number
of simple fractional values of $n$. The results obtained in this section
can also be used to predict the ground-state structures for $1 < n < 2$.
In section~\ref{ngt2}, we consider
the ground-state structures for fillings greater than two. In these
calculations, we use simple geometric arguments to arrive 
at the ground states. This analysis is performed under the assumption that the 
range of the intervortex interaction, which is set by the penetration depth, 
is much smaller than the spacing between the pinning sites. We show that
the ground-state structures obtained from this simple analysis match the
ones obtained from simulated annealing. This analysis is extended to
rectangular and triangular pin lattices in section~\ref{rect}. 
In section~\ref{mag}, the 
zero-temperature equilibrium magnetization of a vortex system in a square array 
of pinning centers is obtained by first calculating the ground-state  
energy as a function of the magnetic induction and then finding the 
applied field from a numerical differentiation of 
the data. The ground-state energies for different values of the magnetic
induction are obtained using a simulated annealing procedure. The calculated
$B-H$ curve exhibits interesting commensurability effects, manifested as
plateaus occurring at simple rational values of the filling fraction $n$.
The main results of our study are summarized in 
section~\ref{summ}.

\section{Ground states for a square pin array with filling ratio less
than one}
\label{nlt1}

 We consider a superconducting film that has a square 
 array of pinning sites with lattice constant $d$. The magnetic field is
 assumed to be perpendicular to the surface of the film.
 The ``matching field'' $B_{\Phi}$ is then given by 
 $B_\phi \equiv \Phi_0/d^2 $, and the filling fraction $n$ is 
 given by $n = B/B_\phi = B d^2/\Phi_0$ where $B$ is the magnetic induction.
 We assume that the pinning potential is much stronger  
 than the intervortex interaction, but is of extremely short range. The 
 large strength of the pinning potential implies that the vortices 
 must occupy pinning sites as long as the number of vortices 
 does not exceed the number of pinning sites. We also assume that a 
 pinning site can not accommodate more than one vortex. If the pinning 
 centers in the film are microholes, then this assumption amounts to  
 the requirement that the radius of each hole is close to two times
 the coherence length $\xi$ of the superconductor. These assumptions ensure
 that interstitial vortices appear only when the filling fraction $n$
 is greater than unity. The assumed short range of the pinning potential
 can be justified if the defect diameter is small compared to the 
 defect spacing $d$. Another assumption that we will make in most of
 our calculations is that the intervortex interaction falls off rapidly
 with distance. This is ensured if 
 the penetration depth $\lambda$ is much smaller that the pin-lattice 
 spacing $d$. In our calculations, we take the ratio $\lambda/d$ to be $10$.
 This value is appropriate for the pin lattice of Ref.\onlinecite{ion}.
 We consider temperatures
 that are low enough to neglect effects of depinning and vortex-lattice
 melting. The problem of finding the structure of the vortex system then
 reduces to locating the ground state in the presence of the pinning
 potential.
  
 Consider now fillings of the form $n=1/q$, $q$ being an integer 
 greater than one.
 Let us look at Bravais lattices that can be formed for a specific $n$ by
 distributing the vortices on the square pin lattice with one vortex per basis.
 The motivation for considering such lattices is that this will
 automatically ensure that there is no shear of the vortices with respect to
 the pin lattice, since the forces on a vortex due to other vortices will
 add up exactly to zero. The unit cell area of these structures has to be 
 $q.d^{2}$. So the possible unit cells can be obtained by factorizing $q$ into 
 products of the form $r.s$ ($r$ and $s$ are integers), 
 arranging the vortices at the corners of  
 rectangles of dimension $rd \times sd$, and then sliding the parallel 
 sides relative 
 to each other. This procedure produces a large number of structures 
 depending on the 
 value of $n$ and we have to pick the one that minimizes the energy. 
 For small values of $q$,
 this can be done by hand, but as $q$ becomes large and highly 
 factorizable, the number of possible structures increases rapidly. 
 In such cases, we have
 resorted to computers to generate these structures and compare their energies.

 The structures so obtained for fillings $1/2$ and $1/4$ match
 those found in the imaging experiment~\cite{ion}.  
 Also for fillings $1/2, 1/3, 1/4, 1/5, 1/8, 1/9, 1/10$ and $1/15$, 
 we find the same structures as those obtained
 by solving the ``greedy lattice gas model''~\cite{glat} exactly. 
 This is understandable because when the intervortex interaction 
 falls off rapidly 
 as the distance is increased above the defect spacing $d$, the ground state
 can be attained by finding the lattice that {\em maximizes the 
 shortest distance
 between vortex pairs}. If two structures have same value and number of shortest
 distances, then the next shortest distance should be maximized, and so on.
 For fillings $1/2, 1/3, 1/4$, and $1/5$, our analysis also yields the same 
 structures as those found in the large $U_{e}$ ($U_{e}$ is 
 the energy of on-site repulsion 
 between two electrons) limit of the neutral 
 Falicov-Kimball model~\cite{fmmodel}.
 In Fig. 1(a,b,c), we have shown the structures 
 so obtained for a few fillings of the 
 form $n = 1/q$. The ground-state structure shown in Fig. 1(b) for $n=1/5$
 is different from that found in Ref.\onlinecite{simul1a} from a simulated
 annealing calculation. This difference is probably due to the use of a 
 different (logarithmic) intervortex potential in Ref.\onlinecite{simul1a}.

 Let us now compare the energies of the nearest and next-nearest neighbors 
 in one of these lattices. The interaction energy between 
 two vortices separated
 by a distance $r$ is given by the expression,
 \begin{eqnarray}
 U(r) = \frac{\Phi_{0}^{2}t}{8\pi^{2}\lambda^{2}}K_{0}
 \left(\frac{r}{\lambda}\right)
 \end{eqnarray}
 where $K_{0}$ is the zeroth order Hankel function of imaginary argument and 
 $t$ is the film thickness. For $n = 1/2$, the nearest neighbor distance is 
 $\sqrt{2}d$ and the next-nearest neighbor distance 
 is $2d$. So the interaction energies 
 are, to within a constant prefactor, given by 
 \begin{eqnarray}
  U_{n} \propto K_{0}\left(\frac{\sqrt[]{2}d}{\lambda}\right) 
  =  0.2 \times 10^{-6}, \nonumber
 \end{eqnarray}
 \begin{eqnarray}
  U_{nn} \propto K_{0}\left(\frac{2d}{\lambda}\right) =  
  0.6 \times 10^{-10}. \nonumber
 \end{eqnarray}
 One can see here that there is orders of magnitude 
 difference in these energies 
 which cannot be compensated by differences arising from interactions with more
 distant neighbors. This difference is going to be more prominent at lower
 densities. This tells us that the maximization of the 
 shortest intervortex distance
 \begin{figure}[htbp]
 \epsfysize=9cm
 \centerline{\epsfbox{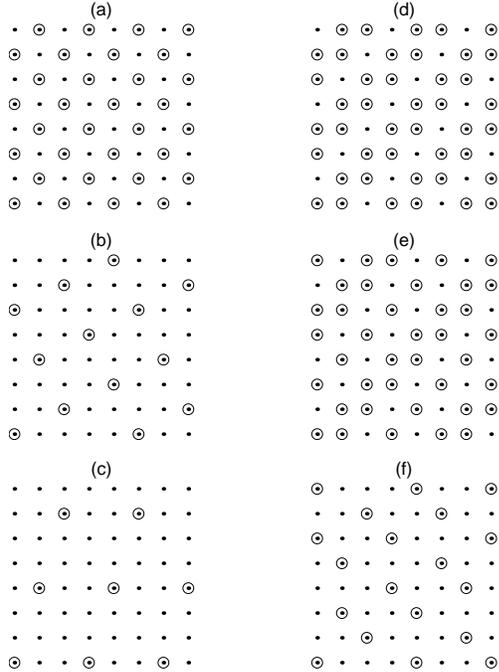}}
 \vspace{0.5cm}
 \caption{\narrowtext The ground state structures for a few filling fractions
 $n < 1$. The different filling fractions are 
 (a) 1/2, (b) 1/5, (c) 1/9, (d) 2/3,
 (e) 3/5, and (f) 2/7. The dots in the figures represent the pinning sites
 and the circles represent the vortices.}
 \end{figure}
 in a lattice for a given filling would lead to the ground states,
 provided the lattice spacing is large compared to the penetration depth of 
 the film. This, in fact, is exactly the definition of the greedy lattice gas. 
 However, one has to be cautious about this method because,  
 as noted in Ref.~\onlinecite{glat}, the structures 
 can be strongly dependent on the 
 form of the potential in certain ranges of $n$ and we can even have aperiodic 
 structures as ground states. The ground-state structures shown here
 have been cross checked with simulations to ensure that they are
 indeed the lowest energy configurations. To give an example of a case where
 this treatment does not lead to the true ground state, we found
 that for filling $1/16$, the energy per vortex for the structure with minimum
 energy obtained this way was greater than that for filling $1/15$, 
 implying that the structure obtained for $n=1/16$ was not the ground state. 

 When the filling fraction is of the form $p/q$ with $p$ not equal to one, 
 one can look for ground states in a subset of structures where the 
 unit cell has size $qd^{2}$ with $p$ vortices in a basis. 
 We have shown in Fig. 1(d,e,f) some of the ground-state structures 
 obtained this way. These structures match those obtained from our simulated 
 annealing calculation. These ground states show 
 the ``stripe'' structure predicted by
 Watson~\cite{glat} and Kennedy~\cite{fmmodel} in appropriate density ranges.

\section{Ground states for a square pin array with filling ratio greater
than two}
\label{ngt2}

 If $n$ is greater that $1$ but smaller than $2$, then the 
 ground-state structures are similar to the ones for the case of $n$ 
 less than one. The only difference is that the pinning sites are all 
 occupied and the centers of the square unit cells of the pin lattice
 now act as new pinning centers for interstitial vortices. 
 But things look different when the filling goes
 above the value $n = 2$. For such values of $n$, we can no longer 
 place the interstitial vortices 
 at the centers of the squares and look for simple structures obtained this
 way.  Also, we now have to start looking into the stability of the 
 structures since the square symmetry would not be present.

\subsection{The ground state for $n = 5/2$}

 Here we are faced with the task 
 of placing more than one vortex in a square. Before going to the 
 problem of finding the ground state for $n = 5/2$, 
 let us ask a more basic question: given a single unit 
 cell of the square pin lattice with each corner occupied by a vortex.
 how can we arrange two more vortices inside this square so as 
 to minimize the energy. Since ``greedy lattice gas'' has been a good 
 approximation for the previous cases, we try to tackle this problem 
 by ``maximizing the shortest distance'' method.  In order to stabilize an 
 interstitial vortex by maximizing the shortest distance, its 
 distance from at least three nearest vortices must be the shortest 
 distance. It is also required that these nearest set of
 vortices must be spread in such a way that if we draw  
 straight lines from the
 vortex in question to these neighbors, the angles formed by adjacent
 lines must be less than $180^{o}$. The proof of this statement is given 
 in Appendix~\ref{app1}. 
 
 It can be seen from the symmetry of the problem that we have to
 place the two vortices on the lines joining the centers of the sides to 
 meet the condition mentioned above. This leaves us with only two possible 
 ways of doing it, which are shown in Fig. 2.
 \begin{figure}[htbp]
 \centerline{
 \epsfysize=4.2cm
 \epsfbox{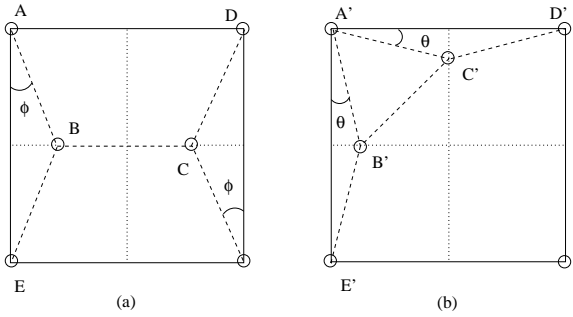}}
 \vspace{0.5cm}
 \caption{\narrowtext Putting two vortices in a square. 
 The configuration in (a)
 is the global energy minimum, and the configuration in (b) can at best be a 
 local minimum of the energy. The angles $\theta = 15^{o}$ 
 and $\phi = 24.49^{o}$. 
 The square is drawn for easy visualization and the dotted lines are the 
 bisectors of the sides. Note that there is already one pinned 
 vortex at each of 
 the corners of the square. The distances $AB = BC = BE = D_{ma}$ 
 and $A'B' = B'C'
 = B'E' = D_{mb}$.}
 \end{figure}
 In (a) the shortest distance $D_{ma}$ can be obtained by solving the equation
 \begin{eqnarray}
 \frac{1}{4} (d - D_{ma})^{2} + \frac{d^{2}}{4} = D_{ma}^{2}. 
 \end{eqnarray}
 On solving this equation, we get $AB = BC = BE = D_{ma} = (\sqrt[]{7}
 - 1)d/3$. In (b) the vortices $A', B'$ and $C'$ form an equilateral triangle.
 Thus the nearest neighbor distance $A'B' = B'C' = B'E'$ is
 \begin{eqnarray} D_{mb} = \sec(15^{o})\frac{d}{2} \end{eqnarray}
 The angle $\phi$ in (a) is $24.49^{o}$, and the angle $\theta$ in 
 (b) is  $15^{o}$. The interaction energies corresponding to these two 
 distances for $d/\lambda = 10$ are
 \begin{eqnarray}
 U_{AB} \propto K_{0}\left(\frac{\sqrt[]{7}-1}{3\lambda}d\right) 
 \simeq 2.2 \times 10^{-3} \nonumber
 \end{eqnarray}
 \begin{eqnarray}
 U_{A'B'} \propto  K_{0}\left(\frac{\sec(15^{o})}{2\lambda}d\right) 
 \simeq 3.0 \times 10^{-3}. \nonumber
 \end{eqnarray}
 From comparing these two energies it is clear that (a) is the global
 minimum, whereas  configuration (b) can at best be a local minimum.
 
 Coming back to the $n = 5/2$ case, we now have to build up the 
 lattice with equal number of two types of squares - one with two 
 interstitial vortices and the other with one interstitial vortex. 
 Note that here we have neglected structures that have three or more 
 interstitial vortices inside a square unit cell because 
 such structures would drastically
 bring down the nearest neighbor distance. Let us now look at the possible 
 units cells of size $2d \times 2d$ that can be made out of these two
 types of squares. These are shown in Fig. 3. 
 \begin{figure}[htbp]
 \epsfysize=6cm
 \centerline{\epsfbox{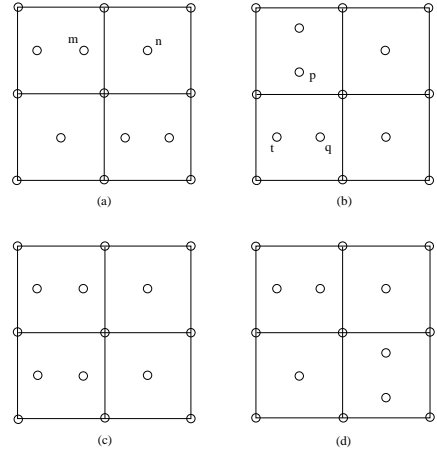}}
 \vspace{0.5cm}
 \caption{\narrowtext{Possible $2d \times 2d$ unit cells for $n = 5/2$}. The
 unit cell (a) is the lowest energy configuration for $d/\lambda$ = 10. For
 much larger values of $d/\lambda$, the unit cell with the lowest energy
 will be the one in (b).}
 \end{figure}
 If one constructs the lattice with these unit cells, 
 configuration (b) offers the least number of next-nearest neighbors, the 
 number of nearest neighbors being the same in all the cases. So one can 
 expect (b) to be the ground state unit cell,
 at least for large values of $d/\lambda$. 
 However, unit cell (a) is preferred if $d/\lambda$ is
 not very large. This can be understood in the following way: the 
 advantage that (b) has over (a) is that it has only half the number of 
 next-nearest neighbors (interactions like that between vortices $m$ and $n$) 
 compared to (a). But this is done at the cost of bringing in 
 interactions like those between vortex pairs ($p,q$) and ($p,t$)
 for every ``gain'' of a next-nearest neighbor interaction. 
 The energies of these two interactions for 
 $d/\lambda = 10$
 are found to be quite close. These energies are
 \begin{eqnarray}
 U_{mn} \propto K_{0}(r_{mn}/\lambda) \simeq 3.2 \times 10^{-4}, \nonumber
 \end{eqnarray}
 \begin{eqnarray}
 U_{pq} \propto K_{0}(r_{pq}/\lambda) \simeq 1.9 \times 10^{-4}, \nonumber
 \end{eqnarray}
 where $r_{mn}$ and $r_{pq}$ are the distances between vortices $m$ and $n$,
 and $p$ and $q$ in Fig.3, respectively. It is clear from this comparison 
 that the unit cell (a) would be preferred for
 $d/\lambda = 10$.
 
 The ratio of interaction energies of the next-nearest and the nearest 
 neighbors is 0.07 for this lattice when $d/\lambda = 10$.  
 This energy difference is appreciable here, so that we can 
 expect that the unit cell we arrived at is the correct one. Note that 
 any net force that might be present on one of the interstitial vortices  
 due to the asymmetry in the structure can be compensated by extremely 
 small displacements from the positions obtained from the ``maximization 
 of the shortest distance'' method.

\subsection{The ground state for $n = 3$}

 When the filling fraction equals $3$, we have to build up the lattice using
 blocks of type (a) in Fig 2. Again, looking at unit cells of size
 $2d \times 2d$ or smaller, we have the configurations shown in Fig. 4
 to consider.
 \begin{figure}[htbp]
 \epsfysize=6cm
 \centerline{\epsfbox{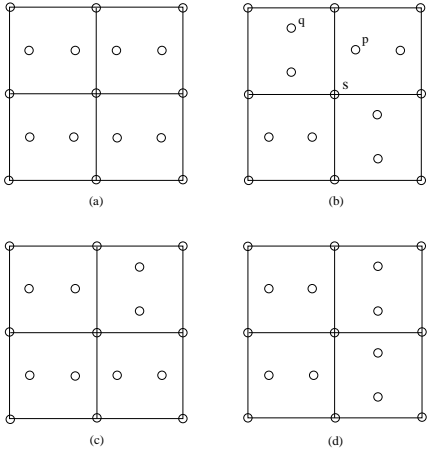}}
 \vspace{0.5cm}
 \caption{\narrowtext{Possible $2d \times 2d$ unit cells for $n = 3$}.
 Configuration (b) can be easily seen to offer the maximum next-nearest
 neighbor distance, the nearest neighbor distances being the same and equal
 in number in each case.}
 \end{figure}
 Here it is easy to see that unit cell (b) is preferred
 over the others. This is because it is the configuration that maximizes the
 minimum distance between any two vortices in different squares, the distances
 between vortices within one square being the same in all the configurations.
 Again comparing the nearest interaction and the next-nearest one, we have
 \begin{eqnarray}
 U_{n} \propto K_{0}\left(r_{sp}/\lambda\right) 
 \simeq 2.2 \times 10^{-3}, \nonumber
 \end{eqnarray}
 \begin{eqnarray}
 U_{nn} \propto K_{0}\left(r_{pq}/\lambda\right) 
 \simeq 1.9 \times 10^{-4}. \nonumber
 \end{eqnarray}
 There is appreciable difference between these two values, and hence, 
 the ground state we have arrived at is reasonable.  
 When the filling lies between $2$ and $3$, one 
 can safely assume that the ground state structure can be built up using
 squares of type (a) in Fig. 2, and squares that have one vortex at the center.
 In fact we make use of this in our simulations to arrive at the ground states, 
 as described in section~\ref{mag}.

\subsection{The ground state for $n = 4$}

 Here we have to place three interstitial vortices in one square. This is 
 nontrivial since even if we ensure that the shortest distance is 
 maximized in one square, two vortices in nearby squares may be 
 closer to each other than the shortest distance within 
 a square when we create the lattice.
 Note that we did not come across this problem in the $n = 5/2$ 
 or $n = 3$ fillings. To illustrate this problem, we show in Fig. 5(a) 
 a single cell of a square lattice with three interstitial vortices, which 
 may very well be a local minimum configuration.
 \begin{figure}[htbp]
 \centerline{
 \epsfysize=4.2cm
 \epsfbox{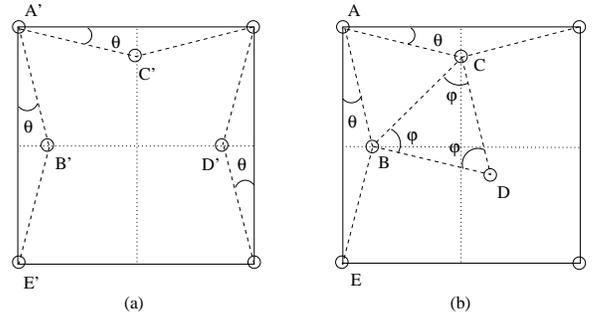}}
 \vspace{0.5cm}
 \caption{\narrowtext Two possible ways of arranging three vortices 
 in a square so as to maximize the shortest distance. The configuration in (a) 
 offers the best arrangement if looked in isolation, but the
 configuration in (b) 
 wins out when one has to construct a lattice of the unit cells. 
 The angles $\theta = 15^{o}$ and $\varphi = 60^{o}$.}
 \end{figure} 
 But if we try to build the lattice using this cell, we can not do it without 
 bringing the vortices in nearby squares closer than the minimum distance in an
 individual cell. So what we need to do is to look for a pattern that will 
 include vortices in different squares while doing the minimization of the 
 shortest distance. We can solve this problem trigonometrically.  
 Consider the figure shown in Fig. 6.  
 The  solution we are looking for can be obtained by solving the equations 
 \begin{eqnarray}
 AB = BC = CA = D_{s},
 \end{eqnarray}
 \begin{eqnarray}
 CB' = AC'' = AP =  D_{s}.
 \end{eqnarray}
 Note that here we have assumed a unit cell size $d \times d$. On solving these 
 equations, we obtain the unit cell shown in Fig. 5(b). In the figure, the 
 angle $\theta$ equals $15^{o}$ and the angle $\varphi$ is $60^{o}$.
 In this lattice, the shortest distance is $D_{s} = \sec(15^{o})d/2$
 and the next shortest distance is $D_{ns} = 
 3d\sec(15^{o})/(4\,\sqrt{2})$.
 
 This simple solution may not be the correct one if 
 the lattice spacing is not large enough. For example, 
 if $d/\lambda = 10$,
 as we have been assuming when comparing energies, then the nearest
 and next-nearest neighbor interaction energies turn out to be really close.
 Hence we can not rule out the possibility of the lattice arranging in
 such a way that the shortest distance is reduced so as to decrease the
 number of nearest or next-nearest neighbors. The relevant energies
 for $d/\lambda = 10$ are
 \begin{eqnarray}
 K_{0}\left(\frac{D_{s}}{\lambda}\right) \simeq 3.1\times 10^{-3},  \nonumber \\
 K_{0}\left(\frac{D_{ns}}{\lambda}\right) \simeq 2.1\times 10^{-3}. \nonumber 
 \end{eqnarray}
 So the ground state obtained above is guaranteed to be the correct one only 
 for much larger values of $d/\lambda$.
\begin{figure}[htbp]
 \epsfysize=5cm
 \centerline{\epsfbox{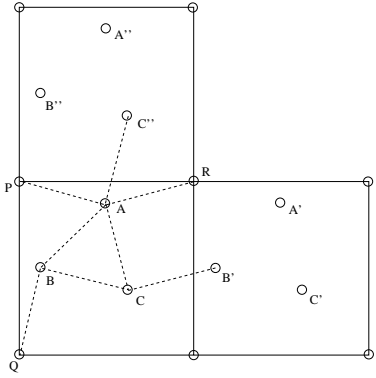}}
 \vspace{0.5cm}
 \caption{\narrowtext The distances one has to equate when maximizing 
 the shortest distance for $n = 4$, taking into consideration distances 
 between vortices in neighboring squares. Note that the unit cell size here
 is one square unit.}
 \end{figure} 
 The ground states we have obtained for $n = 5/2$ and $n = 3$ match with
 the images from experiments~\cite{ion} and also the results of 
 simulations~\cite{simul1}.
 But for $n = 4$, the structures found in experiments and simulations are
 different from the one shown in Figs 5(b) and 6. 
 This is expected, since in the experiment
 the value of the ratio $d/\lambda$ was close to ten. Our simulation with
 $d/\lambda =10$ gives the ground-state structure show in Fig. 7 (a), which
 is similar to the one obtained in the experiment~\cite{ion}.  
 In our simulations with very large values of $d/\lambda$, we get 
 structures similar to that in Fig. 6. 
 The simulation result for $d/\lambda = 50$ 
 is shown in Fig. 7 (b), which matches well with the predicted 
 structure. It is to be noted that the simulation result was obtained 
 by starting the system near the expected ground state. So the claim is that 
 it offers at least a local minimum of the interaction energy. 
 The simulations were carried out for different
 system sizes form $2d \times 2d$ to $10d \times 10d$ to rule out any 
 dependence of the results on the boundary condition.
 
\begin{figure}[htbp]
 \epsfysize=4cm
 \centerline{\epsfbox{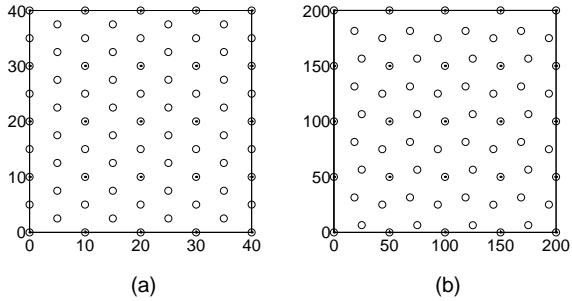}}
 \vspace{0.5cm} 
 \caption{\narrowtext The ground state for $n = 4$, obtained  by simulated 
 annealing when the ratio of the penetration depth to the pin-lattice 
 spacing is 
 (a) $10$ and (b) $50$. The dark dots 
 denote the pinning sites and the axis labels 
 are in units of the penetration depth.}
 \end{figure} 

\section{Ground states for rectangular and triangular pin arrays}
\label{rect}

 One can extend this type of analysis to pinning arrays with other 
 symmetries for 
 finding the least energy structures for simple filling fractions. 
 Let us first consider the case
 of a rectangular array of pinning sites with a pinning unit cell of 
 dimensions $l \times b$, where
 we take $l$ to be the longer side of the rectangle. 
 We shall consider here only the cases where the filling is greater than one. 
 In the absence of the 
 square symmetry, it is obvious that the ground state structure will 
 depend not only on the 
 penetration depth $\lambda$, but also on the ratio $l/b$. 
 In the following analysis, we 
 shall always assume that $\lambda$ is much smaller than $b$, 
 the shorter side of the basic rectangular 
 pinning cell. When the filling is $2$, for values of $l/b$ less than 
 $\sqrt{3}$, the ground 
 state is one where each interstitial vortex is at the center of 
 the rectangle, since this 
 ensures that the shortest distance is maximized (see Fig. 8(b)).
 \begin{figure}[htbp]
 \epsfysize=3.5cm
 \centerline{\epsfbox{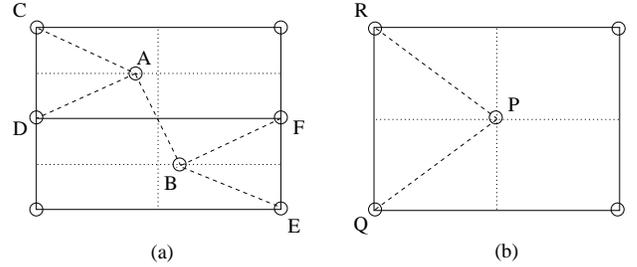}}
 \vspace{0.5cm}
 \caption{\narrowtext The unit cells of the ground state, obtained by
 maximizing the shortest distance, when the filling 
 is $2$ for a rectangular array 
 of pinning sites. (a) The unit cell when the aspect 
 ratio is greater that $\sqrt{3}$. The interstitial vortex is 
 displaced horizontally from the center of the 
 rectangle by distance $D_s$. In the figure, the distances $AB = AC = AD = BF
 = BE$.
 (b) The ground state unit cell when the aspect ratio is less that $\sqrt{3}$, 
 where the interstitial 
 vortex is at the center of the rectangle.}
 \end{figure}
 But when the aspect ratio exceeds $\sqrt{3}$ and the interstitial vortices 
 are placed at the centers 
 of the rectangles, the distance between two interstitial vortices in  
 neighboring cells would be 
 shorter than that between an interstitial vortex and 
 the closest pinned vortex. 
 This would lead to a 
 displacement of the interstitial vortices sideways from the center, 
 along the bisector of the shorter
 sides of the rectangle, to maximize the shortest distance. 
 The resulting structure is shown in 
 Fig. 8 (a). The displacement of the vortex from the center is given by

 \begin{eqnarray}
  D_s = \frac{-l+\sqrt{4l^{2}-9b^{2}}}{6}.
 \end{eqnarray}

 It is worth noting that since the vortex in the center would be moving 
 towards two of the
 pinned vortices, and away from only one interstitial 
 vortex per unit cell,
 the displacement will approach the value given above only 
 when the ratio $l/b$ is 
 appreciably large and in the limit of small penetration depth compared 
 to the sides of the rectangle. 
 For example we have found in our simulations that even when the 
 ratio $l/b$ is $2$, the ground state for 
 $b/\lambda = 15$ is one  where the interstitial vortex is very close to the
 center, whereas for the $l/b= 3$ and  $b/\lambda = 15$, the ground-state
 structure is quite close
 to the one obtained from maximizing the shortest distance. 
 Some of our simulation results for filling 
 equal to two are shown in Fig. 12 (a,b). Also, if the ratio $l/b$ becomes 
 too large, the 
 analysis will have to include more than two of the interstitial vortices, 
 since now the solution like 
 that shown in Fig. 8 (a) can lead to two vortices being closer in the 
 next nearest cells or ones even 
 further apart.

In trying to arrive at the lowest energy structures for fillings $5/2$ and 
$3$, it is important
to determine how one can accommodate two vortices in a rectangular cell 
with the shortest distance being
maximized. 
There are two possible minima that one has to consider: one where the
vortices are arranged along the line dividing the shorter sides,
and one where they are arranged along 
the line dividing the longer sides, as shown in Figs 9 (a) and 9 (b),
respectively. 
The shortest distance in each case is given by,
 \begin{eqnarray}
  D_{s1} = \frac{\sqrt{4l^{2}+3b^{2}}-l}{3}, \\
  D_{s2} = \frac{\sqrt{4b^{2}+3l^{2}}-b}{3}.
 \end{eqnarray} 
 \begin{figure}[htbp]
 \epsfysize=3.5cm
 \centerline{\epsfbox{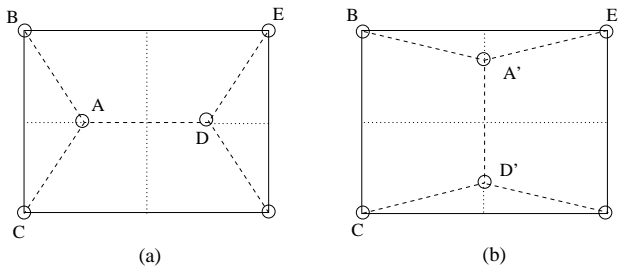}}
 \vspace{0.5cm}
 \caption{\narrowtext The two possible ways by which the 
 shortest distance can be maximized when there
 are two vortices in a rectangular cell. 
 (a) In this case the vortices are placed on the line that
 divides the shorter sides of the rectangle. 
 The distances $AB = AC = AD = DE = D_{s1}$. (b) Here the vortices 
 are placed parallel to the shorter sides and the distances 
 $A'B = A'E = A'D' = CD' = D_{s2}$.
 For $l/b$ less than $2$, the configuration in (b) leads to a larger 
 shortest distance (within
 the cell) than the one in (a). If $l/b$ is greater than $2$, 
 then the distance $D_{s2}$
 becomes greater than $b$ and the vortices spill over into the next cell.}
 \end{figure}
 If one considers the distances within the cell, configuration (b) in Fig. 9 
 gives the lowest energy.
But for large values of the ratio $l/b$, this configuration is disfavored 
since it 
allows the vortices in one rectangle to get close to those in a 
neighboring one. Also, for
 $l/b > 2$, the interstitial vortices ``spill over'' into the next cell, 
 since the distance $D_{s2}$ becomes greater than $b$. So, one has to work
 out the structures for fillings $5/2$ and $3$ case by case. 
 For fillings between $2$ and $3$, one has to choose
 appropriate number of two-vortex rectangles of the right kind  and 
 single-vortex ones and arrange them
 so as to maximize the shortest distance appearing in the structure. 
 We have looked at the $2 \times 2$
 unit cells possible for filling $5/2$ for two values of the aspect 
 ratio, $l/b=4$ and $l/b=5/4$.
 The unit cells that provide the largest minimum distance 
 are shown in Fig. 10. Note that when 
 \begin{figure}[htbp]
 \epsfysize=7cm
 \centerline{\epsfbox{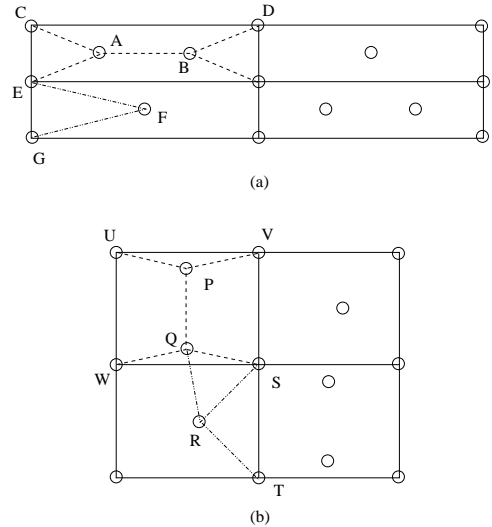}}
 \vspace{0.5cm}
 \caption{\narrowtext The unit cells for $n = 5/2$ for a 
 rectangular array of pinning sites. (a) The
 unit cell when the aspect ratio is $4$. The distances $AB = AC = AE = BD $.
 (b) The unit cell when the aspect ratio is $5/4$. Here $PV = PU = PQ = QW
 = QS = D_{s2}$  
 and $QR = RS =RT$. When this structure is repeated periodically, the 
 ``image'' of the vortex at $P$ would be at the same
 distance $QR$ from $R$. This would ensure the stability of the vortex 
 at $R$.} 
 \end{figure}
 $l/b = 4$, the vortices are arranged parallel to the longer side and 
 in the other case, parallel to the 
 shorter side. Also when $l/b = 5/4$, the vortex in the single-vortex 
 rectangular cell 
 is not located at the center, but slightly displaced sideways 
 along the bisector of the shorter sides
 to facilitate the maximization of the second shortest distance involved. 
 One should again keep in mind
 that this sort of analysis will go wrong if the aspect ratio is too large, 
 since then the distances
 between vortices in next-nearest or further neighbor cells 
 will become important. 
 In Fig. 12 (c) we have shown
 the ground state structure obtained from simulations for aspect 
 ratio $4$, with $b/\lambda = 5$ and in
 Fig. 12 (d), the ground state for aspect ratio $5/4$, with $b/\lambda = 24$.
 In Fig. 12 (d), one can notice the slight shift of the vortex from 
 the center in rectangular units having a
 single interstitial vortex. The shift is not as large as expected from 
 the above analysis. This is due to the large but finite value of $b/\lambda$.
 \begin{figure}[htbp]
 \epsfysize=7cm
 \centerline{\epsfbox{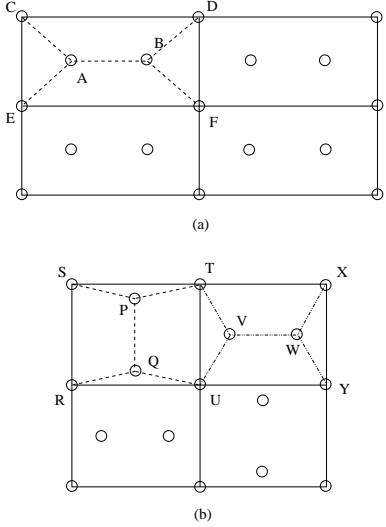}}
 \vspace{0.5cm}
 \caption{\narrowtext Unit cells for $n=3$ for a rectangular array of pins. 
 (a) The unit cell when the aspect ratio is $2$. The distances $AB = AC = 
 AE = BD = BF$. 
 (b) The unit cell for the same filling but for aspect ratio $5/4$. 
 Here the distances
 $PQ = PS = PT = QR = QU = D_{s2}$ and $TV = UV = VW = WY = WX = D_{s1}$.}
 \end{figure}

For filling equal to $3$, the lowest energy structures obtained by considering 
$2 \times 2$ unit cells
for two values of the aspect ratio, $2$ and $5/4$, are shown in Fig. 11. 
Here too, for large values of the
 aspect ratio, the structure is composed of rectangular cells where the 
 interstitial vortices are aligned parallel to 
 the longer side (Fig. 11(a)), whereas when the aspect ratio is smaller, 
 the structure is made up of an alternating arrangement
 of rectangular cells of both types (Fig. 11(b)). 
 The simulated annealing results for similar values of the 
 aspect ratio (Fig. 12 (e,f)) yield the structures obtained from the above
 analysis.

 Ground state structures obtained by simulated annealing for a rectangular
 pin array with $l/b=2$ and integral values of $n$ 
 are reported in Ref.\onlinecite{simul1b}. 
 In that study, the intervortex interaction was assumed to depend
 logarithmically on the intervortex distance. 
 The ground state structure found in Ref.\onlinecite{simul1b}
 for $n=2$ is similar to that shown in Fig. 12(b), but the structure found
 there for $n=3$ is quite different from those of Fig. 12(e,f). This
 is another example of the importance of the detailed nature 
 of the intervortex 
 interaction in determining the structure of the ground state.
 \begin{figure}[htbp]
 \epsfysize=7.5cm
 \centerline{\epsfbox{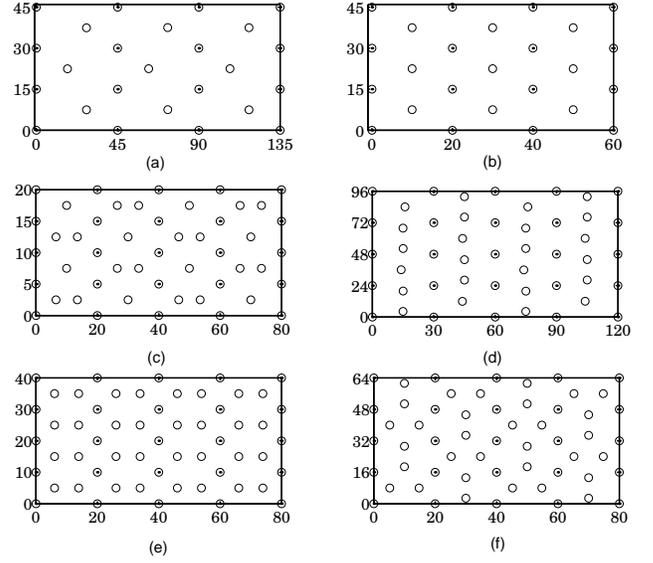}}
 \vspace{0.5cm}
 \caption{\narrowtext  The ground state structures obtained by 
 simulated annealing when the pinning array has rectangular symmetry.
 The parameters for different plots are, (a) $n = 2$, $l/b = 3$ and 
 $b/\lambda = 15$, (b) $n = 2$, $l/b = 4/3$ and $b/\lambda = 15$, 
 (c) $n = 2.5$, 
 $l/b = 4$ and $b/\lambda = 5$, (d) $n = 2.5$, $l/b = 5/4$ and 
 $b/\lambda = 24$,
 (e) $n = 3$, $l/b = 2$ and $b/\lambda = 10$ and (f) $n = 3$, $l/b = 5/4$ and
 $b/\lambda = 16$. The dots denote the pinning centers and the circles 
 represent the vortices.
 Note that the box sizes are not to scale.}
 \end{figure} 

 For a triangular array of pinning sites, it is easy to see that when 
 the filling is greater than $1$ and 
 less than 3, the interstitial vortices will be placed on the centroids 
 of the triangles in the limit where
 one can safely apply the method of maximization of the shortest distance.
  So the ground states when $n$ is between $1$ and $3$
 will be made up of parallelogram cells of the form shown in Fig. 13. 
 These unit cells match well with the results
 of molecular dynamics simulations~\cite {trigs,simul1}.   
 \begin{figure}[htbp]
 \epsfysize=3.1cm
 \centerline{\epsfbox{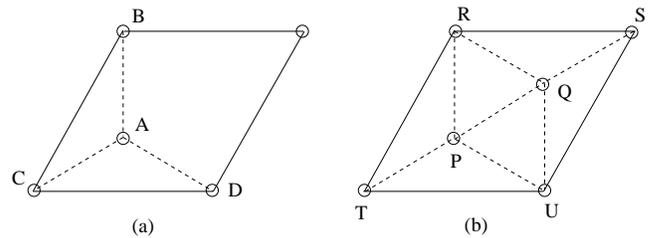}}
 \vspace{0.5cm}
 \caption{\narrowtext Basic building blocks for generating ground states 
 for triangular pinning arrays when the filling
 is between $1$ and $3$. (a) When there is a single interstitial vortex 
 in a parallelogram, the shortest distance
 can be maximized by placing it at one of the two centroids of the 
 triangles involved. The distances $AB = AD = AC = a/\sqrt{3}$,
 where $a$ is the length of the side of a pinning cell. 
 (b) When two vortices are to be placed in a single pin cell, they 
 have to be at the two centroids.} 
 \end{figure}
 
\section{Equilibrium magnetization of the ground states}
\label{mag}

 In this section, we describe a calculation of 
 the zero-temperature, equilibrium magnetization 
 of a thin-film superconductor in the presence of a square array 
 of pinning sites.
 The region in the $B-H$ plain we are interested in is that just 
 above $H_{c1}$, 
 when the flux tubes start entering the sample. The idea is to 
 find the free energy $F$ of the ground state as a function of the magnetic 
 induction $B$, and then to obtain the applied magnetic field $H$ by taking
 a derivative of the free energy with respect to $B$.
 Since we are considering the zero-temperature case, 
 the free energy is just the 
 internal energy of the flux lattice.
 Since we are looking for a nearly continuous variation of the internal
 energy for taking the derivative, we need 
 to locate the ground states for filling fractions separated by small 
 intervals. 
 This would be difficult to do analytically, since the unit cells for some 
 filling fractions can be arbitrarily
 large. Also, as $n$ becomes large, the simple procedure of 
 maximization of the shortest distance 
 is not going to yield the correct ground-state structures. So we have 
 resorted to simulations to determine 
 the ground states. In particular, we have used the simulated annealing 
 technique 
 to locate the global minima (or at least low-lying local minima close 
 in energy to the
 global ones) of the part of the internal energy associated with
 intervortex interactions.
 
 The Helmholtz free energy per unit volume of the superconductor 
 at zero temperature 
 in the presence of the pinning sites is 
 \begin{eqnarray}
 F_{s}(n) = \frac{n\epsilon_{l}}{d^{2}} +E_{n} - \frac{n\epsilon_{p}}{d^{2}}
 \end{eqnarray}
 where the first term is the line energy, 
 the second term is the interaction energy
 and the third term is the pinning energy.  
 Here, $\epsilon_{l}$ is the line energy 
 per unit length, $\epsilon_{p}$ is the pinning energy per unit length and  
 $E_{n}$ 
 is the interaction energy per unit volume for  
 filling fraction $n$. 
 We note here that the pinning energy increases linearly with $n$ till 
 $n$ becomes one and then remains constant, since multiple occupation of a
 pinning center is not allowed.
 Further, for simplicity,  we express the pinning energy as  
 \begin{eqnarray}
 \epsilon_{p} = m\epsilon_{l}
 \end{eqnarray}
 where $m$ is a positive number whose magnitude 
 depends on the nature of pinning. 
 The interaction part of the free energy, $E_{n}$, is 
 the computational input. Once we know the free energy, we can compute 
 the applied magnetic field $H$ using the relation 
 \begin{eqnarray} \frac{\partial F_{s}}{\partial B} = \frac{H}{4\pi}.
 \end{eqnarray}
 Using the standard expression~\cite{tink} for $\epsilon_{l}$ and 
 taking the logarithm of the Ginzburg-Landau number 
 to be equal to 2, we get the following expression for the applied field 
 as a function of the  filling fraction:  
 \begin{eqnarray}
 H = \frac{\Phi_{0}}{2\pi\lambda^{2}}[1-m\{1-\Theta(n-1)\}]+
 \frac{\partial E_{n}^\prime}{\partial n}. 
 \end{eqnarray}
 Here $E_{n}^\prime$ is given by the expression
 \begin{eqnarray}
 E_{n}^\prime = \frac{\Phi_{0}}{2\pi\lambda^{2}N}\sum_{i>j} 
 K_{0}(r_{ij}/\lambda), 
 \end{eqnarray}
where $N$ is the number of basic pinning squares in the system 
and $r_{ij}$ is the separation between vortices $i$ and $j$
in the ground state for the filling fraction $n$. 
 \begin{figure}[htbp]
 \epsfysize=6cm
 \centerline{\epsfbox{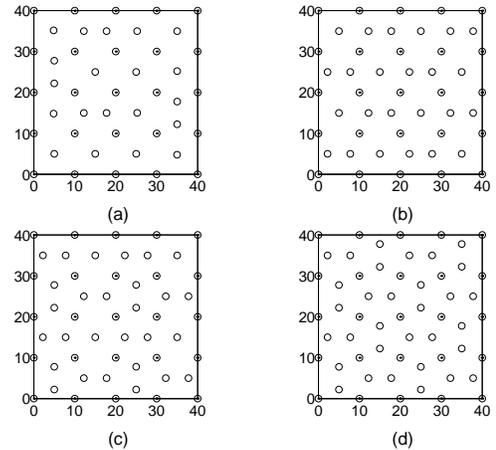}}
 \vspace{0.5cm}
 \caption{\narrowtext The ground states for (a) $n = 9/4$, (b)  $n = 5/2$,
  (c) $n = 11/4$, (d) $n = 3$
 obtained by simulated annealing, as discussed in the text.
 The unit-cell size is $4d \times 4d$.
 The dark dots denote the pinning sites and the circles denote the vortices.
 The axis labels are in units of
 the penetration depth.}
 \end{figure}

The size of the systems we simulated varies from $2d\times 2d$ to $8d\times 8d$.
In all cases, 
we used periodic boundary conditions to minimize surface effects. 
 So the minimum difference between two consecutive 
 filling fractions was $\Delta n = 1/64$. 
The ratio $d/\lambda$
 was taken to be $10$, as in our previous analysis.
 In order to save computation time, the vortices were allowed to stay only at 
 the pinning sites when the filling was less than one. 
 For fillings between 
 one and two, every pinning site was occupied by a vortex 
 which was never moved 
 and the extra ones were allowed to move near the centers of the 
 basic pinning squares. 
 When the filling was greater that two and less than three, 
 the vortex configurations were constructed using
  basic units of squares containing one vortex at its center and squares  
 containing two vortices 
 placed such that the shortest distance within
 a square is maximized (as in Fig. 2(a)). These units were then moved around 
 and twisted while cooling to arrive at the minimum energy states. 
 This procedure helped us to track low-lying minima faster than if we 
 allowed vortices to move freely.   
 Once the basic structure was thus obtained, the vortices were allowed 
 to move freely during a second cooling schedule starting from a lower 
 temperature to obtain the lowest-energy structure.
 In Fig. 14 we have shown some of the ground state 
 structures we have obtained this way for fillings between $2$ and $3$.
 For fillings 
 $5/2$ and $3$, we find that the structures match those obtained in 
 experiments 
 \cite{ion}, as well as in our analysis using maximization of the 
 shortest distance. 
 The structures for $n = 9/4$ and $n = 11/4$ may not be the actual ground 
 states, 
 either due to the smallness of the unit cell of our simulation or due to 
 the presence of many nearly degenerate local minima. 

 \begin{figure}[htbp]
 \epsfysize=5.5cm
 \centerline{\epsfbox{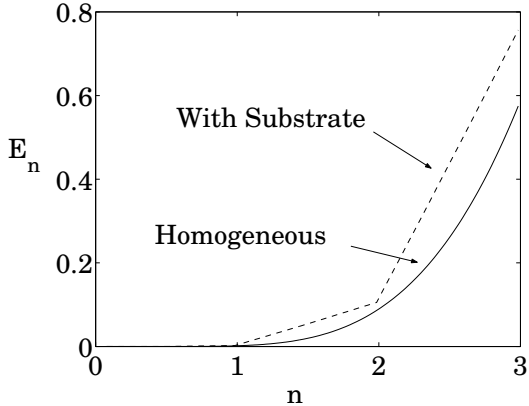}}
 \vspace{0.5cm}
 \caption{ The energy of the vortex lattice in the presence of a
  square array of pinning sites (upper dashed curve). The energy of the
  triangular vortex lattice in the absence of pinning is also shown for
  comparison (lower solid curve).
  The system size is $8d \times 8d$. The abscissa is the filling fraction $n$
  and the ordinate is the total energy per unit thickness in units of 
  $\Phi_{0}^{2}/(8\pi^{2}\lambda^{2})$.}
 \end{figure}

 In Fig. 15 we have plotted the ground state energies obtained from the 
 simulation for
 different fillings. 
 The simulation unit cell was $8d \times 8d$ and the energies were
 computed for fillings $1/64$ to $3$. 
 The upper curve shows the results obtained in the presence of the 
 pinning sites
 and the lower curve is the energy of the triangular lattice for 
 the same density
 of vortices. Note that we have not included the pinning 
 energy in the plot. This would bring down the upper curve below the curve
 for the pin-free case. 

 \begin{figure}[htbp]
 \epsfysize=5.5cm
 \centerline{\epsfbox{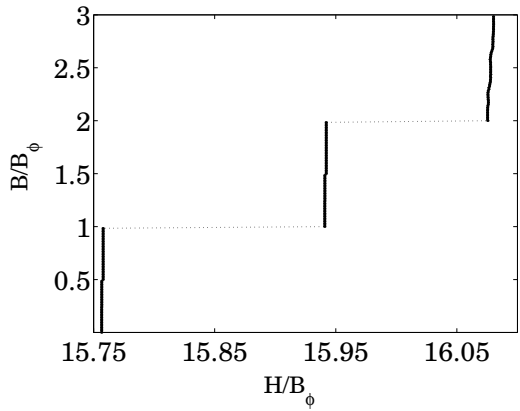}}
 \vspace{0.5cm}
 \caption{The dependence of the magnetic induction $B$ on the applied field 
 $H$ in the entire region
 of our simulation. A number of plateaus can be seen at points corresponding 
 to simple rational filling fractions mentioned in the text. Only the vertical 
 lines correspond to the data points obtained from our calculation; 
 the dotted lines are guides to the eye. Both $B$ and $H$ have been scaled
 by the matching field $B_\phi$.}
 \end{figure}

 From the energy versus filling fraction data, one can find the applied field
 using Eq.(11) and then compute the magnetization $M$ using the relation
 \begin{eqnarray}
 B = H + 4{\pi}M
 \end{eqnarray}
 In Fig. 16 we have plotted $B$ versus $H$ in the entire 
 range of filling for which simulations were carried out, from $n=0$ to $n=3$.
 Figs 17 and 18 show magnified versions of this plot in the regions between
 filling fractions
 $0$ and $1$, and between $2$ and $3$, respectively.
 Note that we have not explicitly included the pinning energy term 
 in our analysis. This term would
 just add a constant contribution to $H$ for fillings
 up to $1$. The features of the curve from $n = 1$ to $n = 2$  are 
 the same as those in the 
 interval between $n = 0$ to $n = 1$. This is  due to the fact that the 
 ground state 
 structures are similar in the two regions (see section \ref{ngt2}). 
 The $B-H$  plot shows flat regions at values of $B$ corresponding 
 to fillings $1/8$, $1/5$, $1/4$, $1/2$, $3/4$, $4/5$, $7/8$, $1$, $9/8$, 
 $6/5$, $5/4$, $3/2$, $7/4$, $9/5$, $2$ in the filling fraction range between 
 $0$ and $2$. 
 Also, in the range of $n$ between $2$ and $3$, there are roughly two plateaus, 
 appearing near $n = 2.3$ and $n = 2.6$. 
 \begin{figure}[htbp]
 \epsfysize=5.5cm
 \centerline{\epsfbox{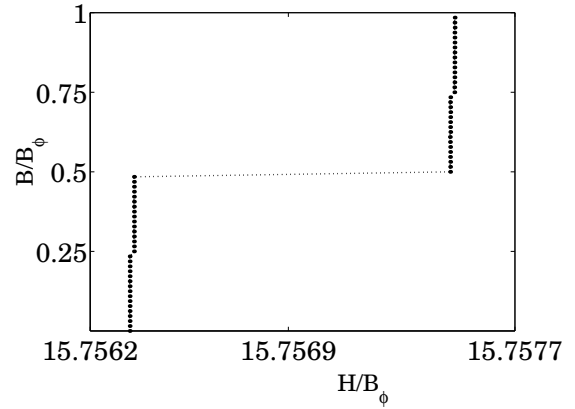}}
 \vspace{0.5cm}
 \caption{Expanded view of the plot of Fig.16 in the region between
 $n=B/B_\phi=0$ and $n = 1$. One can
 see plateaus appearing at $n = 1/4$, $n = 1/2$ and $n = 3/4$. The dark dots
 are the data points. The light dotted line is drawn as a guide to the eye.}
 \end{figure}

 The observed values of the filling fractions between $n=0$ and $n=1$ at 
 which the 
 plateaus occur indicate that these values of $n$ correspond to
 fillings where the introduction of a new vortex into the system leads to the 
 appearance of a shorter distance than those existing in the 
 lattice or in its dual (i.e. the lattice obtained by replacing
 particles by holes, and vice versa)~\cite{glat}.
 This makes sense because the introduction of this shorter 
 distance brings in a larger energy scale, leading to a discontinuous
 change in the derivative of the energy with respect to the filling $n$.
 In the simulations, we have not
 scanned very small intervals of $n$. Also, 
 for some of the fillings,
 the ground state may not have been obtained in our 
 simulated annealing calculation. For these reasons, 
 we can not say anything definite about the true
 nature of the $B-H$ curve. It is possible that this curve has plateaus and
 discontinuities occurring at all scales (e.g. at all rational values of $n$).
 The noisy nature of the $B-H$ plot in the range where $n$ lies between $2$
 and $3$ (see Fig. 18) is also due to these difficulties.
 However,  this plot shows clear signatures of two  plateaus that appear 
 near $n = 2.3$ and $n = 2.6$. These can be understood as happening when 
 first there is an occurrence of squares containing two vortices coming next 
 to each other diagonally (as in Fig 3(a)), and again when 
 they have to be next to each other with a common side (as in Figs 3(b) 
 and 3(c)), 
 as the value of $n$ increases from $2$ to $3$.   
 \begin{figure}[htbp]
 \epsfysize=5.5cm
 \centerline{\epsfbox{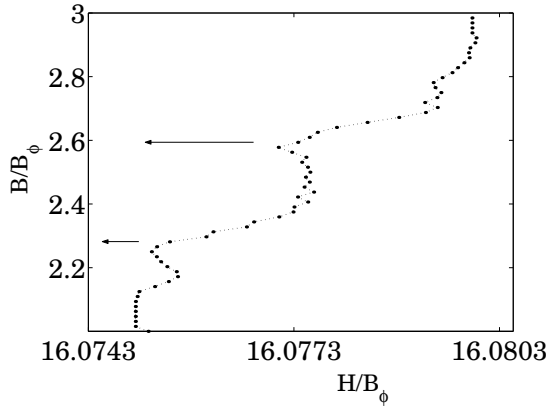}}
 \vspace{0.5cm}
 \caption{Expanded view of the plot of Fig.16 in the region between $n = 2$ 
 and $n = 3$. One can see plateaus at fillings $n \simeq 2.3$ and 
 $n \simeq 2.6$, as indicated by the arrows.
 The dark dots are the actual data points and the light dotted 
 line is shown as a 
 guide to the eye.}
 \end{figure} 
 
 \section {Summary and discussions}
 \label{summ}
 In this paper, we have reported the results of an analytic calculation of
 the lowest-energy states of a vortex system in the presence of a 
 periodic array of strong pinning centers, each of which can trap only one
 vortex. We have considered several different lattice structures of the 
 pin array and a large number of filling fractions in the range
 between zero and four. The analytic calculations are based on the principle
 of maximization of the shortest intervortex distance. We have argued that
 this principle leads to the exact ground states when the spacing of
 the defect lattice is large compared to the range of the intervortex
 interaction set by the value of the penetration depth. This principle has
 been used, in combination with simple geometric considerations, to obtain
 the ground states for several values of the filling fraction $n$. The
 ground-state structures so obtained are found to be identical to those
 found in imaging experiments~\cite{ion} and in earlier 
 simulations~\cite{simul1,trigs}. We have also carried out simulated annealing
 calculations of the ground states in order to test some of the predictions
 of the analytic approach. In all cases, we found that the analytic results
 agree with those of our numerical calculation.

 We have also described the results of a numerical calculation of the
 equilibrium 
 magnetic induction $B$ and magnetization $M$ of a planar superconductor
 with a square array of pinning centers as functions of the externally
 applied field $H$. We show that the interplay between the lattice spacing 
 of the pin array
 and the intervortex separation set by the value of $B$ leads to 
 interesting commensurability effects, showing up as plateaus and
 discontinuities in the $B$ vs $H$ plot at simple rational values of the
 filling fraction $n$. Anomalies in the {\it irreversible} magnetization of
 thin-film superconductors with periodic arrays of pinning centers have been
 observed at certain integral values of $n$ in experiments~\cite{exp1,exp2} and
 simulations~\cite{simul1}. The presence of a periodic array of pins is
 also expected~\cite{th1} to produce anomalies in the equilibrium
 magnetization of the high-temperature vortex liquid at small integral
 values of $n$. Our results show that these commensurability effects are 
 more pronounced in the field-dependence of the equilibrium magnetization and 
 magnetic induction of such systems
 in the low-temperature vortex-solid regime. Experimental investigations of
 these effects would be most welcome.
\appendix
\section{Condition for maximizing the shortest distance}
\label{app1} 
 Given a particle fixed at some point in a plane, the problem is how to
 place three other particles around the first one in such
 a way that if we try to move the first particle from its position, it will
 get closer to at least one of the three particles.
 The solution is as follows. If we draw straight lines from the particle
 that we want to move to the other particles, then each angle between 
 adjacent lines must be less than $180^{o}$. In other words,
 it should not be possible to draw a straight line through the particle 
 in question in such a way that all the other three particles lie on one 
 side of the line.
 
 Let the particle that we want to move be at $P_{0}$ (see Fig. 19). 
 Let us place two particles at
 $P_{1}$ and $P_{2}$, anywhere on the plane. 
 \begin{figure}[htbp]
 \epsfysize=6cm
 \centerline{\epsfbox{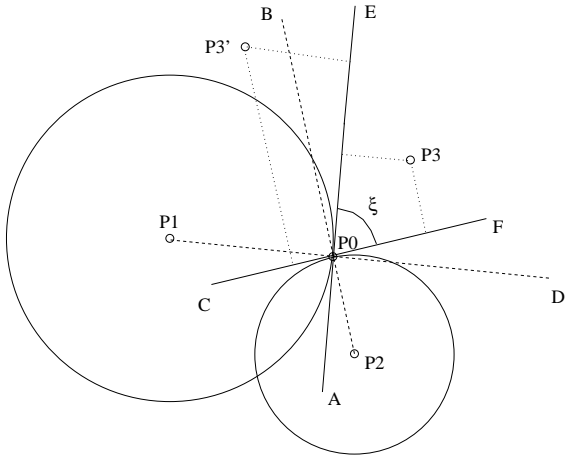}}
 \vspace{0.5cm}
 \caption{Geometry of the problem of maximizing the shortest distance 
 (see text).}
 \end{figure}
 This can be done since one of the angles between any two lines
 will always be less than or equal to $180^{o}$. 
 In Fig. 19, $AE$ and
 $CF$ are the tangents to the circles centered at $P_{1}$ 
 and $P_{2}$ and passing through the point $P_{0}$. 
 The presence of these particles
 restricts to within the angle $\xi$ the direction in which 
 the particle at $P_{0}$ can move 
 without decreasing
 its distance to $P_1$ and $P_2$.
 Now let us check where
 we can fix the third particle such that the aforementioned condition is met,
 that is the particle at $P_{0}$ cannot be moved without bringing it closer to
 one of the particles at $P_{1}$, $P_{2}$ and $P_{3}$. We will consider two 
 possible cases - the third particle in the region $BP_{0}D$ of the
 plane and outside this region.
 
 When the third particle is in the region $BP_{0}D$ (point $P_{3}$ in the
 figure), it is not possible for the particle $P_{0}$ to move from its 
 position without decreasing the distance to any one of the particles. 
 This can be seen if one drops the perpendiculars to the lines $AE$ and $CF$ 
 from the point $P_{3}$. If the point $P_{0}$ is moved along either of these 
 lines, it will be getting closer to the point $P_{3}$.
 But that in turn implies that it cannot be moved into the region 
 $FP_{0}E$ at all without decreasing the distance from the point $P_3$. 
 
 If the particle is outside the region $BP_{0}D$, for example a point like
 $P_{3}'$, then it is easy to see that by moving along one of the 
 tangents to the circles at the point $P_{0}$, the particle at $P_{0}$ can 
 move away from all the three other points. Thus we find that the earlier 
 statement we made is proved.  This gives us a nice way to maximize the 
 shortest distance since all one has to do is to make the three distances 
 involved equal, so that if the central particles tries to move, then one of 
 the distances has to decrease.    


\end{document}